\documentclass[singlecolumn,notitlepage,groupedaddress,11pt]{revtex4-1}

\usepackage{subfig}
\usepackage{units}
\usepackage{amsmath}
\usepackage{graphicx}% Include figure files
\usepackage[hidelinks]{hyperref}

\begin{document}

\title{Realistic Shortcuts to Adiabaticity in Optical Transfer}

\author{Gal Ness, Constantine Shkedrov, Yanay Florshaim, Yoav Sagi}

\affiliation{Physics Department, Technion -- Israel Institute of Technology, Haifa 32000, Israel}
\date{\today}
\begin{abstract}
Shortcuts to adiabaticity (STA) are techniques allowing rapid variation of the system Hamiltonian without inducing excess heating. Fast optical transfer of atoms between different locations is a prime example of an STA application. We show that the boundary conditions on the atomic position, which are imposed to find the STA trajectory, lead to highly non-practical boundary conditions for the optical trap. Our experimental results demonstrate that, as a result, previously suggested STA trajectories generally do not perform well. We develop and demonstrate two complementary methods that solve the boundary conditions problem and allow the construction of realistic and flexible STA movements. Our technique can also account for non-harmonic terms in the confining potential.   
\end{abstract}

\maketitle

\section{Introduction\label{sec:Introduction}}

Future quantum technologies will require deep understanding and exquisite control of the underlying system dynamics. The Hamiltonian would have to be changed as fast as possible while still reaching a specific target state. However, rapid variation of the Hamiltonian might introduce undesired excitations and heating.
\emph{Shortcuts to Adiabaticity} (STA) refers to a class of solutions to this problem, all based on cleverly tailoring the fast driving protocol such that the system reaches a desired final state which is adiabatically connected to the initial one, but without adhering to the adiabatic condition \cite{STA_review}. It  can be roughly divided into two subcategories: \emph{counterdiabatic} and \emph{invariant-based} drivings. Counterdiabatic methods continuously suppress excitations by utilizing a time-dependent auxiliary term in the Hamiltonian, and are therefore harder to implement experimentally. Invariant-based methods, on which we focus in this work, take the system back to the adiabatic path only at the end of the process by engineering the trajectory to satisfy the boundary conditions at initial and final times and, in general, the Ermakov equation in between \cite{Chen2011}. Among the many fields who can benefit from STA are quantum computation \cite{Sarandy2011,Santos2015,Santos2016,Stefanatos2018}, quantum control \cite{Bason2011,Takahashi2017,Mortensen2018,Levy_2018}, quantum thermodynamics \cite{Campo2011,Li2018} and transport \cite{Reichle2006,DGO_optimal_transport,torrontegui2011fast,Bowler2012,Fuerst2014,Sels2018,Corgier2018} or manipulation  \cite{Bulatov1998,Chen2010,An2016,Deng2018,Zhou2018} of ions and ultracold neutral atoms.

The rich toolbox available in ultracold atomic experiments makes these systems particularly attractive as platforms for quantum simulation and quantum computation  \cite{Bloch2008,Saffman2010,Gross2017}. In particular, the interaction of atoms with a far-off-resonance laser light introduces negligible dissipation and can be effectively described as a conservative potential for the atoms \cite{Grimm200095,Schlosser2001,Lester2015}, with potential depth that is proportional to the laser intensity. Ultracold atoms can be trapped in the vicinity of potential minimum, which for the lowest energy optical mode, namely a Gaussian beam, is at the focal point (``waist''). Rapid changes may be desired in the trap shape \cite{Bulatov1998,Campo2011,Chen2010,Deng2018} or position  \cite{DGO_optimal_transport,torrontegui2011fast,Fuerst2014,Guery-Odelin2014,Sels2018}. Optical transfer of ultracold atoms can be used to move atoms between different sites or implement quantum gates \cite{Hayes2007,Isenhower2010,Saffman_2016}. Shorter transfer duration is advantageous considering the finite coherence time of any experimental system, but at the same time, it might have an adverse effect on the fidelity of single operations \cite{Saffman_2016}. STA techniques applied to optical transfer can alleviate this conflict.

In a practical implementation of invariant-based STA, a problem arises: the boundary conditions are given for the atomic trajectory while the experimental control is over the position of the trap, which in turn is derived from the atoms position through the equation of motion. This leads to boundary conditions for the trap which may be very difficult to satisfy in reality. For example, it may require the trap and atoms to start at the same position, with the trap having an initial velocity while the atoms are at rest. Here we study experimentally non-adiabatic optical transfer with ultracold atoms and find that indeed this issue almost always harms the performance of known STA trajectories. We show that in order to lift conflicting boundary conditions, it is necessary to increase the number of degrees of freedom in the trajectory. We describe and demonstrate two complementary approaches to achieve this: One, by an addition of a ``correction'' spectral component at the trapping frequency with a specific phase and amplitude. This method can be applied to any existing trajectory. Two, by using a polynomial trajectory with high enough order and incorporating the boundary conditions by a correct choice of the coefficients.

The structure of this paper is as follows: In section \ref{sec:STA in transport problems} we briefly introduce the STA formalism, the required boundary conditions and the use of the final sloshing mode as a measure of undesired excitations added by the movement. In section \ref{sec:The-experimental-system}, we introduce the apparatus, and describe the experimental sequence and probing technique. We then present in section \ref{sec:Experimental-tests-of} our experimental study of several known STA trajectories for optical transfer. Our results show that almost always the movement results in considerable heating. This leads us in section \ref{sec:new_approaches} to develop and demonstrate two new methods to construct proper STA trajectories. We conclude and give our outlook in section \ref{sec:Discussion}.

\section{Shortcut to adiabaticity in transport problems\label{sec:STA in transport problems}}

The optical dipole potential induced by a Gaussian beam can be written as \cite{Grimm200095}:
\begin{equation}
U\left(z,r\right)=-\frac{U_{0}}{1+\left(\frac{z-z_{\cup}}{z_{R}}\right)^{2}}\exp\left\{ -2\frac{r^{2}}{\sigma^{2}}\left[1+\left(\frac{z-z_{\cup}}{z_{R}}\right)^{2}\right]^{-1}\right\} \ \ ,
\end{equation}
where $z$ denotes the atoms coordinate in the laboratory frame of reference and $z_{\cup}$ the position of the potential minimum, both of which will later become time-dependent, $r$ is the radial distance from the beam path, $U_{0}$ is the maximal potential depth, and $z_{R}=\pi\sigma^2/\lambda$ is the Rayleigh range, with $\sigma$ the waist radius and $\lambda$ the wavelength. We expand the potential to fourth order in powers of $z-z_{\cup}$ which leads to the following Hamiltonian (up to a constant):
\begin{eqnarray}\qquad
H=\frac{m}{2}\dot{z}^{2}
+\frac{m \omega_{0}^{2}}{2} \left(z-z_{\cup}\right)^2
	\left(1-2\frac{r^2}{\sigma^2}\right) e^{-{2 r^2}/{\sigma ^2}}
	 \nonumber\\ \qquad \qquad
	-\frac{m \omega_0^2}{2  z_R^2}\left(z-z_{\cup}\right)^4 
	\left(1-4\frac{r^2}{\sigma^2}+2\frac{r^4}{\sigma^4}\right)e^{-{2 r^2}/{\sigma ^2}} \;, \label{eq:Gaussian_U}
\end{eqnarray}
where $\omega_{0}=\left(2{U}_{0}/m{z}_{R}^{2}\right)^{1/2}$ is the harmonic axial trapping frequency. In our experiment we induce motion along the axial direction and in addition the radial trapping frequency is much higher than the axial one ($\omega_r/\omega_z\approx90$). Hence, we average out the radial motion and retain only the axial dependence in the Hamiltonian. This averaging is done over a period of $\omega_r^{-1}$ for which the axial movement is negligible, and the terms $
\left(1-2\frac{r^2}{\sigma^2}\right) e^{-{2 r^2}/{\sigma ^2}}$ and $
\left(1-4\frac{r^2}{\sigma^2}+2\frac{r^4}{\sigma^4}\right) e^{-{2 r^2}/{\sigma ^2}}$ in Eq.\,\eqref{eq:Gaussian_U} are replaced by their ensemble averaged values.
At low temperatures the atoms lie very close to the potential minimum and ${\left\langle r^2\right\rangle}/{\sigma^2} \ll 1$, where $\langle \cdot \rangle$ denotes the ensemble average. Thus, we expand the exponent, retaining terms up to first order, and write the Hamiltonian as:
\begin{equation}
H=\frac{m}{2}\dot{z}^{2} + 
\frac{1}{2} m \omega_{0}^{2} \left(z-z_{\cup}\right)^{2}
\left[1-4\frac{\left\langle r^{2}\right\rangle}{\sigma^2}-\frac{\left\langle \left(z-z_{\cup}\right)^{2}\right\rangle}{z_R^2}\right] \;, \label{eq:harmonic_with_corrections}
\end{equation}
where we also averaged out the fourth order of the axial displacement $z-z_{\cup}$, as ${\left\langle \left(z-z_{\cup}\right)^{2}\right\rangle}/{z_R^2} \ll 1$ for this temperature regime.

Since the temperature in our experiments is very low, in most of what follows we will assume that the two last terms in the rightmost parentheses can be omitted, and we are left with a harmonic Hamiltonian in the axial direction:
\begin{equation}
H\left(t\right)=\frac{m}{2}\dot{z}^{2}\left(t\right)+\frac{m}{2}\omega_{0}^{2}\left[z\left(t\right)-z_{\cup}\left(t\right)\right]^2\;.\label{eq:harmonic_H}
\end{equation}
However, as we show in section \ref{subsec:Anharmonic-effects}, at higher temperatures the non-Harmonic terms become important, and we account for them when constructing the STA trajectories.

We impose the following boundary conditions on the atoms motion:
\begin{subequations}
\label{eq:BC_as_one}
\begin{eqnarray}
z\left(0\right)=0\;,\quad\dot{z}\left(0\right)=0\;,\quad\;,\ddot{z}\left(0\right)=0\label{eq:BC_0}\\
z\left(t_{f}\right)=d\;,\quad\dot{z}\left(t_{f}\right)=0\;,\quad\ddot{z}\left(t_{f}\right)=0\;.\label{eq:BC_t_f}
\end{eqnarray}
\end{subequations}
That is, the atomic movement of $d$ meters during $t_f$ seconds begins and ends with zero velocity and acceleration. As was shown by Lewis and Riesenfeld \cite{Lewis_1969}, with these boundary conditions, and for a general class of Hamiltonians that includes the one in Eq.\,\eqref{eq:harmonic_H}, there exists a dynamical invariant $I\left(t\right)$ that satisfies:
\begin{equation}
\frac{\mathrm{d}}{\mathrm{d}t}I\left(t\right)=\frac{\partial}{\partial t}I\left(t\right)-\frac{i}{\hbar}\left[I\left(t\right),H\left(t\right)\right]=0\; ,
\end{equation}
which can be written explicitly in the harmonic case as:
\begin{equation}
I\left(t\right)=\frac{m}{2}\left\{ \frac{\mathrm{d}}{\mathrm{d}t}\left[z\left(t\right)-z_{\cup}\left(t\right)\right]\right\} ^{2}+\frac{m}{2}\omega_{0}^{2}\left[z\left(t\right)-z_{\cup}\left(t\right)\right]^{2}\;.
\end{equation}
The third equalities in Eq.\,\eqref{eq:BC_0} and in Eq.\,\eqref{eq:BC_t_f} guarantee that  $\left[I\left(0\right),H\left(0\right)\right]=\left[I\left(t_{f}\right),H\left(t_{f}\right)\right]=0$. The state which we would like to conserve can therefore be written at $t=0$ as a superposition of $I$ eigenstates, and since $I$ is invariant under the motion, their  state at $t_f$ is the same as at $t=0$. Hence, under the harmonic approximation, a finite trajectory $z\left(t\right)$ that respects the boundary conditions Eq.\,\eqref{eq:BC_as_one} is a proper STA that will end in an adiabatically connected state. Devising an STA trajectory amounts to finding a non-adiabatic path that satisfies Eq.\,\eqref{eq:BC_as_one}. Several such trajectories were suggested in the literature \cite{Murphy2009, torrontegui2011fast, Chen2011a, Torrontegui2012, Fuerst2014, Guery-Odelin2014, Li2017, Chen2018,Sels2018} and implemented experimentally \cite{Reichle2006,DGO_optimal_transport,Corgier2018}.

However, the boundary conditions Eq.\,\eqref{eq:BC_as_one} are given for the atomic position, while experimentally the control is over the trap position. The latter can be derived from the former using the equation of motion: 
\begin{eqnarray}
m\ddot{z}\left(t\right)=-m\omega_{0}^{2}\left[z\left(t\right)-z_{\cup}\left(t\right)\right]
\quad\Rightarrow \quad z_{\cup}\left(t\right)=z\left(t\right)+\ddot{z}\left(t\right)/\omega_{0}^{2}\;.\label{eq:f_equals_ma}
\end{eqnarray}
The boundary conditions Eq.\,\eqref{eq:BC_0} imply that $z_{\cup}\left(0\right)=0$ and $\dot{z}_{\cup}\left(0\right)=\dddot{z}\left(0\right)/\omega_{0}^{2}$. Similarly, we get that $z_{\cup}\left(t_f\right)=d$ and $\dot{z}_{\cup}\left(t_f\right)=\dddot{z}\left(t_f\right)/\omega_{0}^{2}$. Since $\dddot{z}\left(0\right)$ and $\dddot{z}\left(t_f\right)$ are generally not zero, they require initial and final non-zero velocities for the trap. The first condition is very hard to implement in most physical realizations, as it requires that the atoms are initially at rest and that their position coincide with the trap minimum but there is a finite initial velocity to the trap. In most cases, both the trap and the atoms are at rest before the transport commences. The second condition means that even if the atoms are brought to the adiabatically connected state at the end of the motion, since the trap is still moving it will soon induce atoms excitation. Evidently, we need to require that the trap will also be at rest at the beginning and at end of the transport. This implies:
\begin{equation}
{
\left\{ \begin{array}{@{\kern2.5pt}lc}
z_{\cup}\left(0\right)=0\;,\;\;z_{\cup}\left(t_{f}\right)=d\\
\dot{z}_{\cup}\left(0\right)=\dot{z}_{\cup}\left(t_{f}\right)=0
\end{array}\right.\quad\Rightarrow \quad}
\dddot{z}\left(0\right)=\dddot{z}\left(t_{f}\right)=0\;.\label{eq:new_BC}
\end{equation}
Most STA trajectories do not generally satisfy this extra set of boundary conditions for $\dddot{z}\left(t\right)$ and, as we later demonstrate, their experimental implementation might leave excess energy in the cloud

\begin{figure}
	\centering
	\subfloat[]{\includegraphics{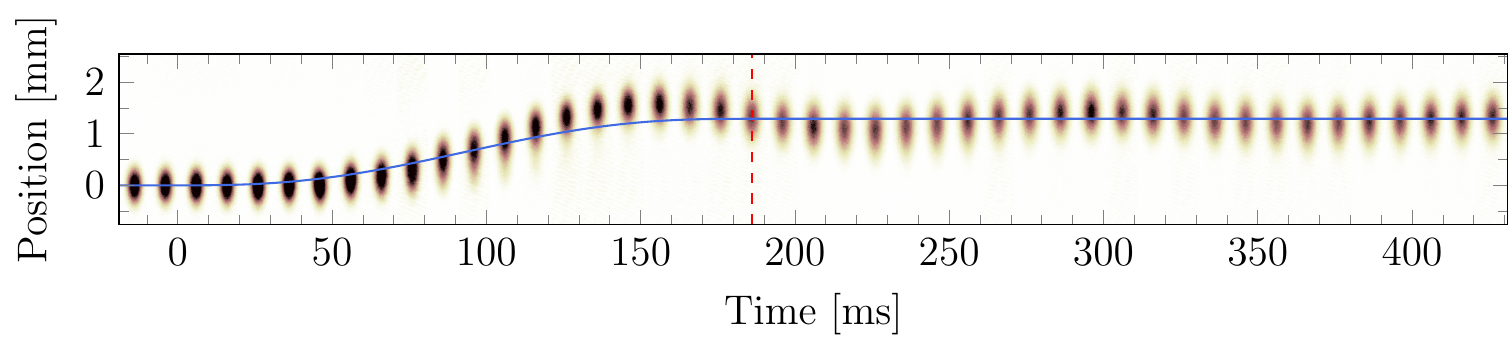}}
	
	\subfloat[]{\includegraphics{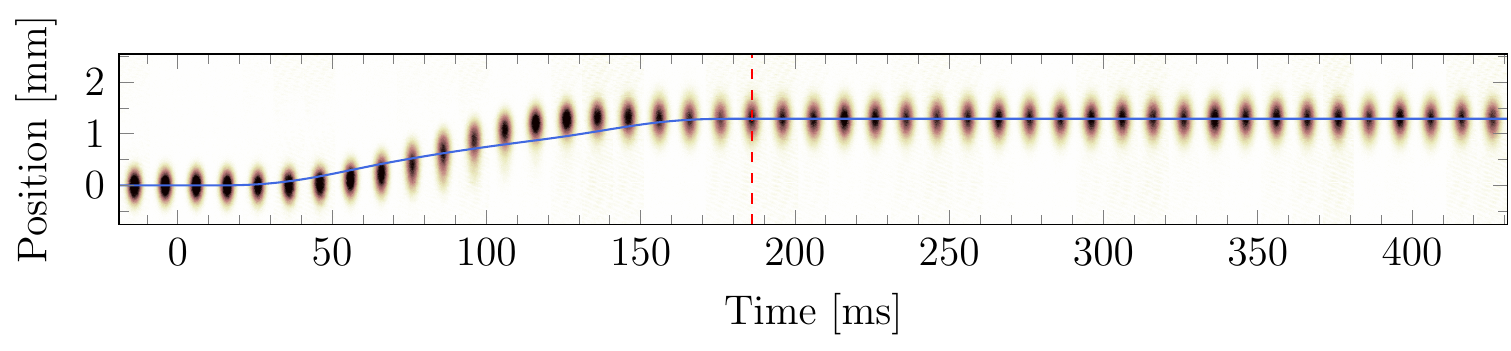}}
	\caption{
		Sequences of absorption images capturing a non-adiabatic transport originally suggested in \cite{torrontegui2011fast} (a) and a corrected version of the same transport (b). The significant sloshing mode exhibited in (a) is suppressed in (b) and by this the associated excess energy is avoided.
		Each panel is composed of $44$ atomic density distributions captured by absorption imaging at different times along and after the translation sequence. Darker shade stands for higher density. The blue curves denote the trap position and the red dashed vertical lines mark the motion end. The trap in (a) was driven along a polynomial trajectory Eq.\,\eqref{eq:poly_traj} of $1.29\thinspace\unit{mm}$ transport during $t_{f}=186\thinspace\unit{ms}$ in a trap of axial frequency $\omega_{0}=2\pi\cdot7.16\left(15\right)\thinspace\unit{Hz}$.
		In (b) the suggested harmonic spectral correction Eq.\,\eqref{eq:harmonic_corrected} was implemented with $\omega_{0}=2\pi\cdot7.11\thinspace\unit{Hz}$, $\phi_{0}=279^{\circ}$ and $A_{0}=105\thinspace\unit{\mu m}$, resulting in $2\left(2\right)\thinspace\unit{\mu m}$ sloshing amplitude compared to $206\left(4\right)\thinspace\unit{\mu m}$ in (a).
		\label{fig:sloshing_capturing}}
\end{figure}

The only motion excited mode in a harmonic potential is the center-of-mass oscillation (sloshing mode). Non-harmonic terms in the Hamiltonian and interactions between the atoms couple between the sloshing mode and higher order modes which eventually lead to increase of temperature. 
Hence, the sloshing mode and temperature are the observables one is required to measure when characterizing realistic implementations of STA in optical transport. As an example, we present in Fig.\,\ref{fig:sloshing_capturing} a series of absorption images of atomic clouds taken during and following a non-adiabatic transport. The upper panel presents the result of a trajectory originally suggested in \cite{torrontegui2011fast}. Considerable sloshing is apparent after the end of the trap motion. This is because the trajectory does not respect the third boundary condition given in Eq.\,\eqref{eq:new_BC}. In contrast, the lower panel shows the same trajectory with our spectral correction method applied. The final state shows no detectable sloshing and minimal increase in temperature.

In the harmonic case, the sloshing mode amplitude can be calculated by integrating over the response function  \cite{baruh2014applied}:
\begin{equation}
\mathcal{A}\left(t_{f}\right)
\equiv\sqrt{ \left[z\left(t_{f}\right)-d\right]^{2}+\frac{\dot{z}^{2}\left(t_{f}\right)}{\omega_{0}^{2}}}
=\left|\intop_{0}^{t_{f}}\exp\left(-i\omega_{0}t\right)\dot{z}_{\cup}\left(t\right)\mathrm{d}t\right|\;.\label{eq:classical_A}
\end{equation}
This is merely the Fourier component at the trap frequency of the trap velocity trajectory. For realistic trap trajectories that maintain Eq.\,\eqref{eq:new_BC} and Eq.\,\eqref{eq:BC_0}, the conditions in Eq.\,\eqref{eq:BC_t_f} translate into zero ultimate sloshing amplitude. This can be used as a guideline for constructing STA trajectories: they should result in zero final sloshing.

\section{The experimental apparatus\label{sec:The-experimental-system}}

\begin{figure}
	\centering
	\includegraphics[width=\columnwidth]{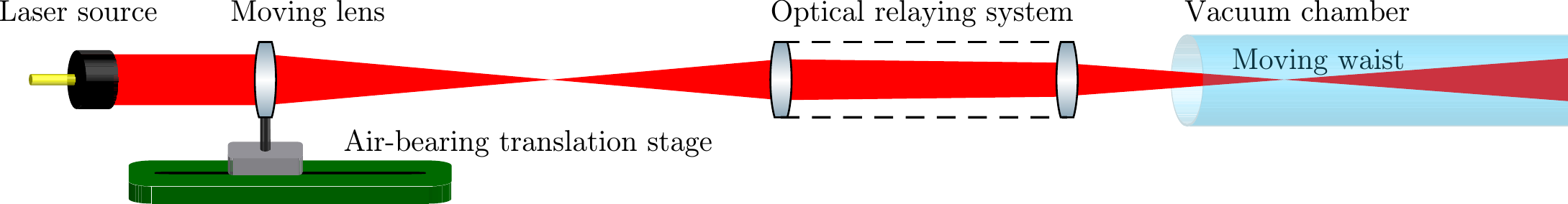}
	\caption{Schematic of the experimental system.
		The Gaussian focal point created by the translatable lens is relayed by the optical relaying system, to transport the atoms held in the focal point inside the ultra-high vacuum chamber. The lens is moved by Aerotech's ABL1500 air bearing stage, specified to sub-micron accuracy and repeatability.
		\label{fig:Experimental_apparatus}}
\end{figure}

The system is composed of three interconnected vacuum chambers. In the first chamber, a two-dimensional magneto-optical trap (MOT) \cite{dieckmann1998two} generates a stream of cold fermionic potassium atoms that fly through a narrow nozzle to the second chamber. There, the atoms are captured and cooled in a three-dimensional dark SPOT MOT \cite{ketterle1993high} on the $\mathrm{D}_2$ line and get farther cooled using gray molasses on the $\mathrm{D}_1$ line \cite{salomon2014gray}. Then, we optically pump the atoms and load them into a QUIC magnetic trap \cite{esslinger1998bose}, where we perform forced microwave evaporation. Next, around $25\cdot {10}^{6}$ atoms at $T/T_{F}\approx4.5$, with $T_{F}$ the Fermi temperature, are loaded into a far-off-resonance optical dipole trap of $\lambda=1064\thinspace\unit{nm}$. This trap is made of a single Gaussian beam, with a waist of $39\thinspace\unit{\mu m}$ and power of about $2.5\thinspace\unit{W}$. The atoms are then transported \cite{Gustavson2001} in approximately a second to the third chamber. This is done by moving a single lens which is a part of an optical relay system that creates the optical trap. The actual movement is performed with an air-bearing translation stage, to reduce vibrations and heating of the ensemble, as depicted in Fig.\,\ref{fig:Experimental_apparatus}. Upon arrival at the third chamber, the waist of the optical trap is $19.45\thinspace\unit{\mu m}$. Forced optical evaporation concludes the preparation stage. In this evaporation we decrease the laser power to some minimal value and then ramp it up back to its final value. By this, we increase the ratio between the trap depth and temperature which effectively narrows the atomic cloud extent with respect to the waist and Rayleigh range. At this final point, the laser power is typically $39\thinspace\unit{mW}$ resulting in a heating rate of about $7\thinspace\unit{nK/s}$. The conditions for the experiments reported here are $N\approx400,000$ atoms at a temperature of $T\approx300\thinspace\unit{nK}$. The atoms are in a balanced mixture of the $m=-9/2,-7/2$ Zeeman states in the $F=9/2$ hyperfine level, which in the applied uniform magnetic field of $\sim185\thinspace\unit{G}$ are weakly interacting with an s-wave scattering length of $\sim236\thinspace a_0$, with $a_0$ being the Bohr radius. Typical trap oscillation frequencies are $\omega_r=2\pi\times646\left(7\right)\thinspace\unit{Hz}$ and $\omega_0=2\pi\times7.16\left(15\right)\thinspace\unit{Hz}$ in the radial and axial directions, respectively. The typical trap Fermi energy is $E_F\approx h\times15\thinspace\unit{kHz}$.

\begin{figure}
	\centering
	\includegraphics{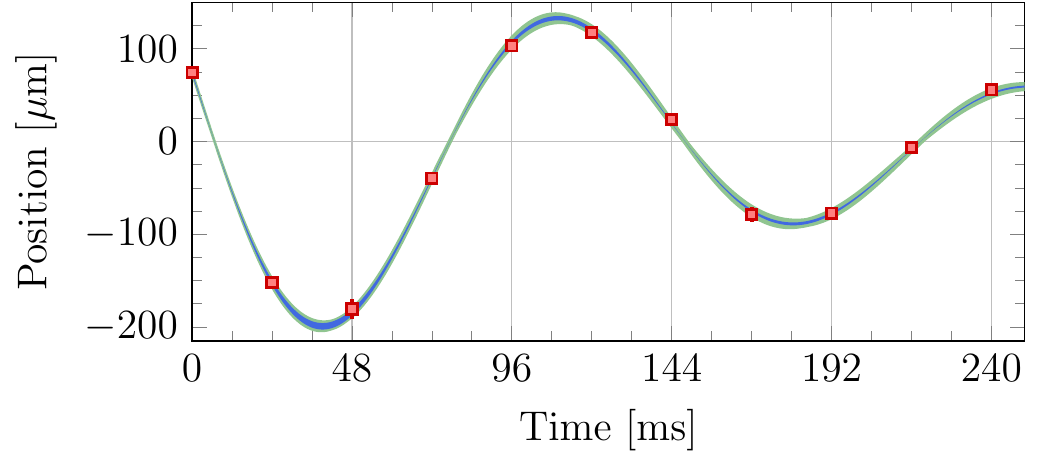}
	\caption{
			Center-of-mass position of the atomic cloud after a ballistic expansion time of $t_{e} = 12\thinspace\unit{ms}$ versus the waiting time after the trap has stopped. Red errorbars are the standard deviation between three independent measurements. The blue curve is a fit to a decaying sine: $\mathcal{A}\sqrt{\left(1-\frac{t_{e}}{\tau}\right)^{2}+\left(\omega_{0}t_{e}\right)^{2}}\sin\left[\omega_{0}t+\varphi+\arctan\left(\frac{\omega_{0}t_{e}}{1-t_{e}/\tau}\right)\right]e^{\left(-t/\tau\right)}$, with a sloshing amplitude $\mathcal{A} = 206\left(4\right)\thinspace\unit{\mu m}$, decay time $\tau = 175\left(6\right)\thinspace\unit{ms}$, $\omega_{0} = 2\pi\cdot7.08\left(3\right)\thinspace\unit{Hz}$ and $\varphi = 133.0\left(8\right)^{\circ}$. The width of the blue curve designates the extracted sloshing amplitude $68\%$ certainty ($1\sigma$) that will be presented as the amplitude error in  following graphs, and the width of its green background represents the error in the other fitting parameters.
	\label{fig:sloshing_fitting}}
\end{figure}

After the cloud reaches equilibrium conditions with a negligible sloshing of a sub-micron amplitude, we execute the STA trajectory by moving again the lens mounted on the air-bearing stage. The stage specifications guarantee that the position of the trap minimum is accurate to within  $1\thinspace \unit{\mu m}$, and the trapping frequencies are constant to within $1.5\%$ during the movement. At each point along the translation we can stop and record the sloshing mode. This is done by waiting for some duration, then abruptly shutting off the trap, letting the atoms expand ballistically and then recording the atomic density distribution using absorption imaging. From these images we can extract the number of atoms, the center-of-mass position and $T/T_F$. The sloshing mode can be reconstructed by fitting a series of such images taken at different waiting times (see Fig.\,\ref{fig:sloshing_capturing}) with a decaying sine. A typical dataset together with the fit is shown in Fig.\,\ref{fig:sloshing_fitting}, that depicts the resemblance of the center-of-mass motion to a decaying oscillation. The typical decay time is $150-200\thinspace\unit{ms}$ due to the spread in the effective harmonic oscillation frequency in Eq.\,\eqref{eq:harmonic_with_corrections} and due to anharmonic terms in the Hamiltonian. The sloshing measurements reported in this work were obtained from eleven different durations, for each we averaged over three repetitions. Note that the time-of-flight expansion before imaging acts effectively as a magnification of the sloshing mode amplitude relative to \emph{in situ}. For the $12\thinspace \unit{ms}$ expansion employed here, the magnification is $\times1.09$.

\section{Experimental test of three uncorrected STA trajectories\label{sec:Experimental-tests-of}}

\begin{figure}
	\centering
	\includegraphics{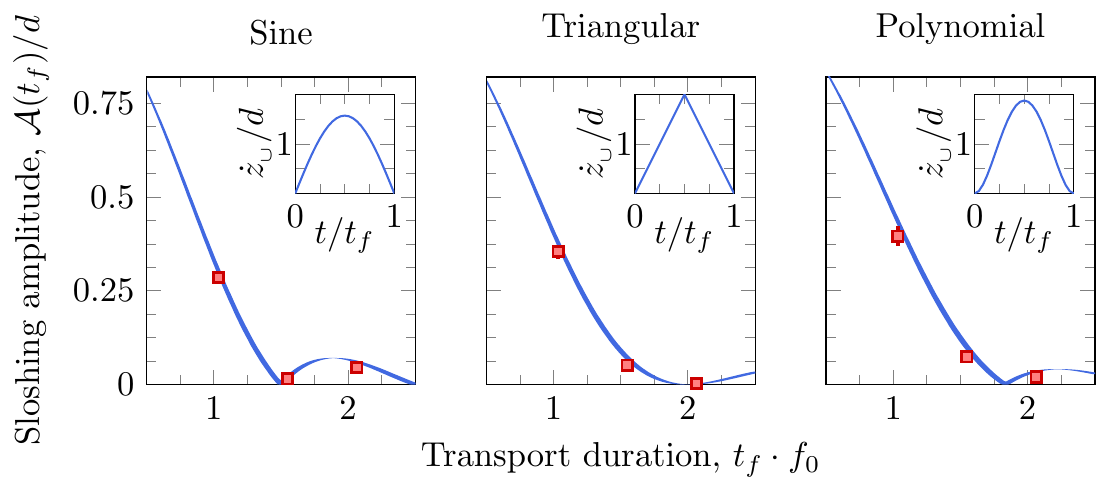}
	\caption{
		Sloshing amplitudes for three non-adiabatic motion trajectories. Red errorbars are the sloshing amplitudes measured following a $1.29\thinspace\unit{mm}$ transport in a trap of axial frequency $f_0=7.32\thinspace\left(8\right)\thinspace\unit{Hz}$. Blue ribbons are harmonic calculations according to Eq.\,\eqref{eq:classical_A} and the width is due to the frequency $f_0$ uncertainty. Each point in the graph corresponds to a different transport duration $t_f$. Inset: time domain representations of the trap velocity profiles.
		% data from 11,12,13mar18.
		\label{fig:sway_types}}
\end{figure}

We have implemented three non-adiabatic trajectories that maintain Eq.\,\eqref{eq:BC_0} and Eq.\,\eqref{eq:new_BC} but not necessarily Eq.\,\eqref{eq:BC_t_f}. Their velocity profiles are depicted in the insets of Fig.\,\ref{fig:sway_types}.
The first trajectory is a \emph{Sine} with a velocity profile given by $\dot{z}_\cup\left(t\right)=\frac{\pi d}{2 t_f}\sin\left(\frac{\pi t}{t_f}\right)$. It satisfies Eq.\,\eqref{eq:BC_t_f} only for $t_f\cdot f_0=\frac{1}{2}+n$ where $f_0=\omega_0/2\pi$ and $n$ is an integer $n\ge 1$. The second trajectory has a \emph{Triangular} velocity profile with a constant acceleration and deceleration given by $4 d/t_f^2$ \cite{DGO_optimal_transport}. It satisfies Eq.\,\eqref{eq:BC_t_f} only for $t_f\cdot f_0=2\cdot n$ with integer $n\ge 1$. The third trajectory is a \emph{Polynomial}, which generally can be written for the atomic position as $z\left(t\right)=d\sum_{n=1}^{\infty}a_{n}\left(\frac{t}{t_{f}}\right)^{n}$. The lowest order polynomial to respect the boundary conditions in Eq.\,\eqref{eq:BC_as_one} is given by \cite{torrontegui2011fast}:  
\begin{equation}
z\left(t\right)=d \left[10\left(\frac{t}{t_{f}}\right)^{3}-15\left(\frac{t}{t_{f}}\right)^{4}+6\left(\frac{t}{t_{f}}\right)^{5}\right]\; .\label{eq:poly_traj}
\end{equation}
In the experiment, we used the polynomial for the trap position instead of the atoms, thus satisfying Eq.\,\eqref{eq:BC_0} and Eq.\,\eqref{eq:new_BC}. However, similar to the other two trajectories, it satisfies Eq.\,\eqref{eq:BC_t_f} only for a discrete set of points which can be calculated numerically, and the first four values are $t_f\cdot f_0\approx 1.835, 2.895, 3.923, 4.938$.

Each of these trajectories was executed for a total movement of $d=1.29\thinspace\unit{mm}$. The resulting sloshing amplitudes $\mathcal{A}\left(t_{f}\right)$ for three non-adiabatic durations $t_f$ are depicted in Fig.\,\ref{fig:sway_types}. For reference, we also plot the calculated sloshing amplitude as given by Eq.\,\eqref{eq:classical_A}. There is a satisfactory agreement between the experiment and theoretical calculations. As expected, zero sloshing is obtained only at specific $t_f\cdot f_0$ values. This places a strict constraint on potential applications of STA. Moreover, it puts a lower limit on the duration of the trajectory which is on the order of $f_0^{-1}$. In what follows we develop two methods to construct STA trajectories that satisfy all eight boundary conditions. In principle, for an ideal harmonic potential these trajectories can be constructed for any desired duration above the fundamental limits \cite{Shanahan2018}. A more practical limit on the shortest possible trajectory stems from the finite trap depth: a faster movement coherently drives the population during the motion via higher energy levels. 

\section{Realistic and flexible non-adiabatic trajectories\label{sec:new_approaches}}
For a trajectory to be \emph{realistic} we require that \emph{both} the trap and atoms will start and end at rest. As we explained earlier, this implies boundary conditions Eq.\,\eqref{eq:BC_as_one} and Eq.\,\eqref{eq:new_BC}. For a trajectory to be \emph{flexible}, we require that it could be constructed for a wide range of distances $d$ and durations $t_f$.
% In the following section we will devise and demonstrate two approaches that...

\subsection{Method I: spectral correction}
\subsubsection{Harmonic potential \label{subsec:Harmonic-correction}}

The idea in this method is to introduce a new spectral component to an existing trajectory which maintains Eq.\,\eqref{eq:new_BC} such that it will satisfy all required boundary conditions and the sloshing amplitude Eq.\,\eqref{eq:classical_A} at $t_f$ will vanish. Since the sloshing amplitude is given by the Fourier component at $\omega_0$ of the velocity profile, we can cancel it by adding a new spectral component at this frequency such that  their sum is zero. This method is very general as it does not assume anything regarding the original trajectory other than it satisfies Eq.\,\eqref{eq:new_BC} and starts with atoms at rest Eq.\,\eqref{eq:BC_0}.

\begin{figure}
	\centering
	\subfloat[]{\includegraphics{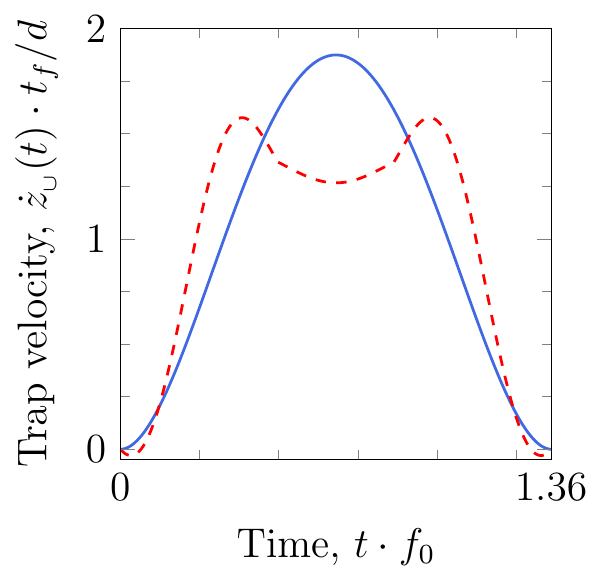}}
	\qquad
	\subfloat[]{\includegraphics{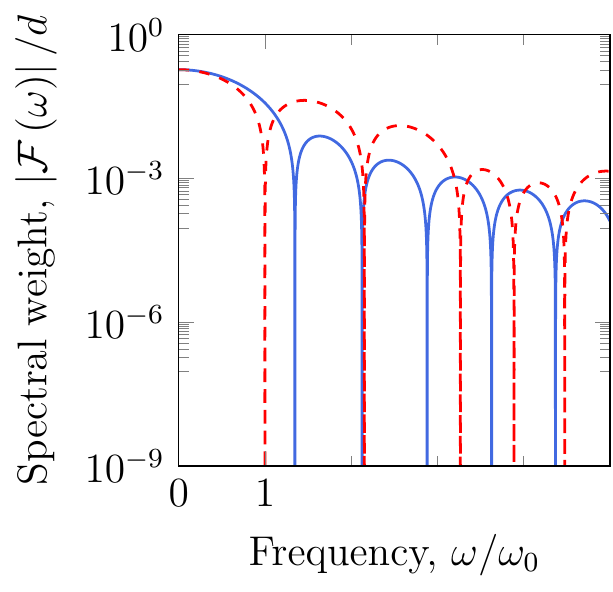}}
	\caption{
		Trap velocity trajectory as a function of time (a) and frequency (b). In solid blue line is the polynomial velocity profile given by Eq.\,\eqref{eq:poly_traj}, and the dashed red line is our corrected trajectory given by Eq.\,\eqref{eq:harmonic_corrected}. As can be seen on the right panel, the Fourier component at the trap oscillation frequency essentially vanishes for the corrected trajectory. By virtue of Eq.\,\eqref{eq:classical_A}, this guarantees zero final sloshing amplitude.}
	\label{fig:Trap_velocity_trajectory}
\end{figure}

Let us denote the original trajectory by $\tilde{z}_\cup$, then the ``corrected'' trajectory is given by:
\begin{equation}
z_{\cup}\left(t\right)=\tilde{z}_\cup\left(t\right)+A\left(t\right)\sin\left(\omega_{0}t+\phi_{0}\right)\;,\label{eq:harmonic_corrected}
\end{equation}
with $A\left(t\right)$ and $\phi_0$ being the amplitude and phase of the correction term, respectively. Our method thus added two new degrees of freedom: the amplitude and phase of the new spectral component. If we plug Eq.\,\eqref{eq:harmonic_corrected} into the boundary conditions, we get that both $A\left(t\right)$ and its first derivative needs to vanish at the starting and ending points. We have chosen to ramp up the correction amplitude as $A\left(t<0.5f_0^{-1}\right)=A_{0}\sin^2\left(\frac{\omega_{0}}{2}t\right)$, and its time-reversed version when decelerating towards the end. Other choices are also possible. According to Eq.\,\eqref{eq:f_equals_ma} and Eq.\,\eqref{eq:new_BC}, the atomic trajectory then satisfies:
\begin{subequations}
\begin{eqnarray}
z\left(0\right)+\omega_{0}^{-2}\ddot{z}\left(0\right)=0\;,\quad
\dot{z}\left(0\right)+\omega_{0}^{-2}\dddot{z}\left(0\right)=0\;,
\label{eq:BC_0_c}\\
 z\left(t_{f}\right)+\omega_{0}^{-2}\ddot{z}\left(t_{f}\right)=d\;,\quad
\dot{z}\left(t_{f}\right)+\omega_{0}^{-2}\dddot{z}\left(t_{f}\right)=0\;.
\label{eq:BC_t_f_c}
\end{eqnarray}
\end{subequations}
The boundary conditions Eq.\,\eqref{eq:BC_0_c} are automatically fulfilled when there is no initial sloshing, as Eq.\,\eqref{eq:BC_0}. There are two more conditions that the atomic trajectory needs to fulfill given by Eq.\,\eqref{eq:BC_t_f}, namely, each of the additives in Eq.\,\eqref{eq:BC_t_f_c} should vanish separately. The two degrees of freedom we have added, i.e. the correction amplitude ($A_0$) and phase ($\phi_{0}$), are then used to suppress the amplitude in Eq.\,\eqref{eq:classical_A}. This can be either done empirically by tuning the parameters and minimizing the resulting sloshing amplitude or numerically by solving the optimization problem of minimizing $\mathcal{A}\left(t_{f}\right)$ in Eq.\,\eqref{eq:classical_A}. As an example, we plot in Fig.\,\ref{fig:Trap_velocity_trajectory} the polynomial velocity trajectory of Eq.\,\eqref{eq:poly_traj} with (dashed line) and without (solid line) our spectral correction. As can be clearly seen in the frequency domain, in the corrected trajectory the spectral component at $\omega_0$ can be suppressed below any desired level.

\begin{figure}
	\centering
	\subfloat[]{
		\includegraphics{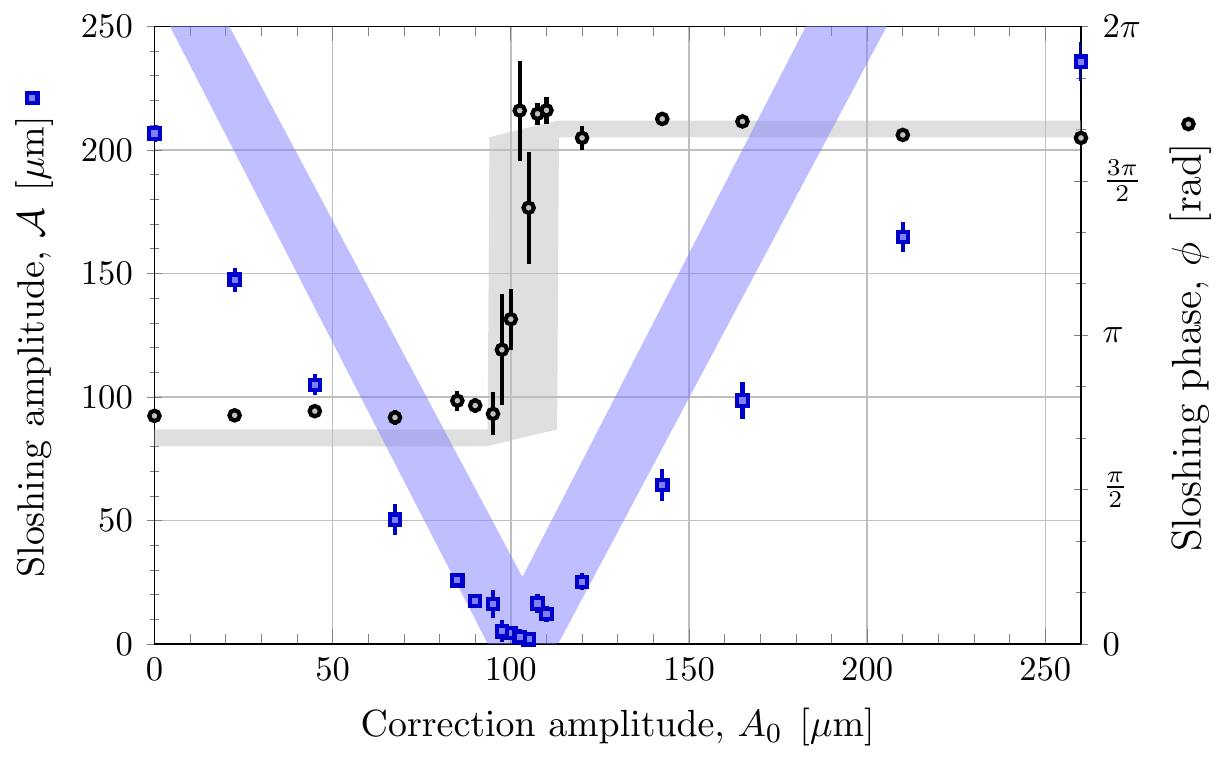} 	
	}

	\subfloat[]{
		\includegraphics{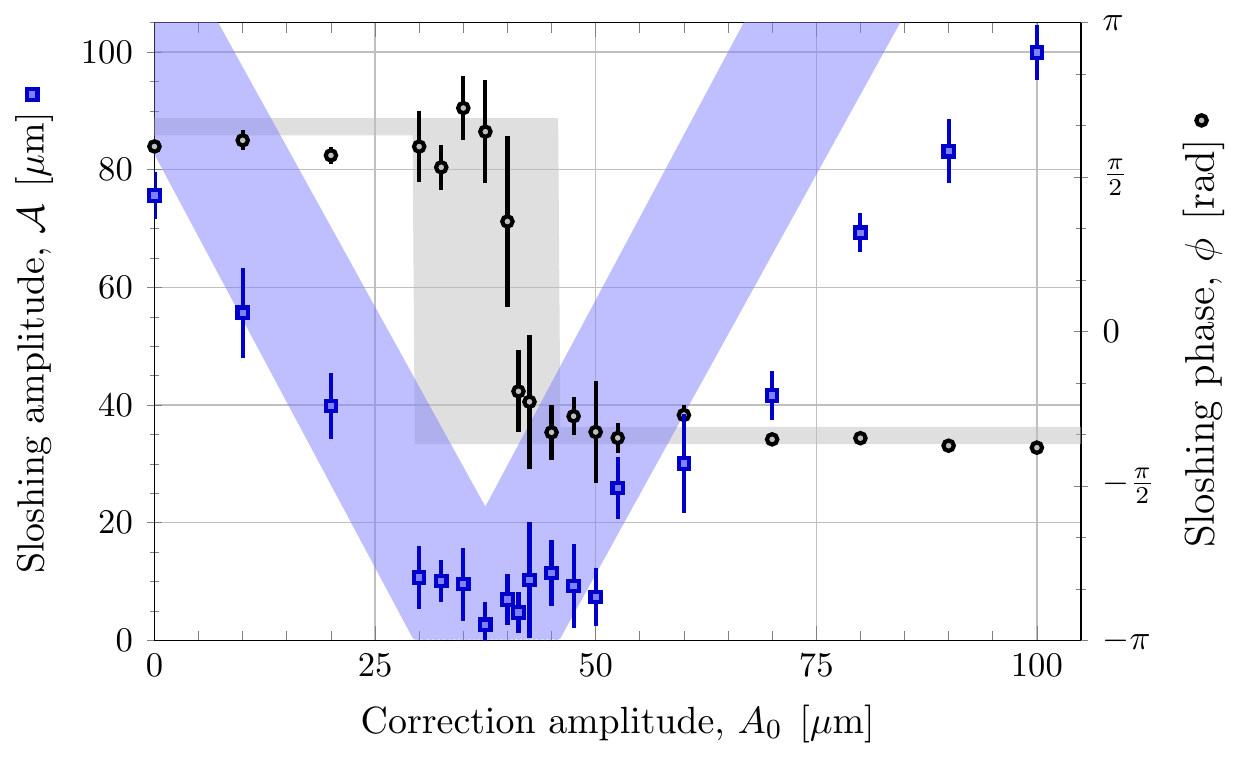}}
	\caption{
		Sloshing mode amplitude (blue squares) and phase (gray circles) versus the harmonic correction amplitude, after non-adiabatic movement with polynomial (a) and sinusoidal (b) trajectories. Theoretical harmonic calculations appear as ribbons with matching colors, where the width is determined by the trapping frequency uncertainty. These measurements were performed following transport of $d=1.29\thinspace\unit{mm}$ lasting $t_{f}=186\thinspace\unit{ms}$ in a trap of axial frequency $\omega_{0}=2\pi\cdot7.16\left(15\right)\thinspace\unit{Hz}$ ($f_{0}\cdot t_{f}=1.33\left(3\right)$).
		% data from 15,16mar17, ribbons freqs are 7.1622(pm0.1469) and 7.1862(pm0.15) (muga and sine) from the fitting results of each experiment.(weighted mean of 20-120 um amplitudes)
		Correction parameters for these measurements are $\omega_{0}=2\pi\cdot7.11\thinspace\unit{Hz}$ frequency and optimal correction phase of $\phi_{0}=279^{\circ},261^{\circ}$ for the polynomial and sinusoidal profiles respectively.
		The expected correction phase, according to minimization of Eq.\,\eqref{eq:classical_A}, is $\phi_{0}=302^{\circ}$ for both cases.
		\label{fig:harmonic_correction}}
\end{figure}

We have tested experimentally our method with polynomial and sinusoidal trajectories. To obtain the minimal sloshing amplitude, we scanned the correction parameters around the calculated optimal values. The measured sloshing amplitudes (blue squares) and phases (gray circles) are plotted in Fig.\,\ref{fig:harmonic_correction} for the polynomial (upper panel) and sine (lower panel) trajectories. A clear minimum in the sloshing amplitude can be observed in both cases. At the optimal correction amplitude, the measured excess energy in the sloshing mode due to the non-adiabatic trajectory is consistent with zero. In contrast, the uncorrected trajectories ($A_{0}=0$) display a substantial sloshing. The atomic cloud dynamics in the non-corrected and corrected ($A_{0}=105\thinspace\unit{\mu m}$) polynomial trajectories are presented in Fig.\,\ref{fig:sloshing_capturing} (a) and (b) respectively. When measuring the temperature after the center-of-mass motion has ceased, we do find an increase of about $200\thinspace\unit{nK}$ for all non-adiabatic transports (corrected and uncorrected). This is probably due to high frequency errors in the actual executed motion that couple through the anharmonic terms to higher vibrational modes of the cloud. We also measure an additional increase in temperature due to the excess energy in uncorrected trajectories compared to corrected ones.

Theoretical calculations shown as ribbons with matching colors in Fig.\,\ref{fig:harmonic_correction} agree reasonably well with the experiments at low sloshing amplitudes, but at higher values they deviate, probably due to contributions from non-harmonic terms in the potential. For both types of trajectories, the sloshing mode phase jumps sharply by $\pi$ when crossing the optimal correction amplitude, as expected from an over-compensated driven harmonic oscillator. We find that the measured optimal correction phase deviates from the theoretical calculation by $23^\circ$ and $41^\circ$ for the polynomial and sinusoidal trajectories, respectively. This is most likely due to experimental imperfections in the execution of the trajectory, to which the phase is most sensitive. This exemplifies another advantage of our method: it can correct for experimental imperfections easily by parameters tuning.  

\subsubsection{Anharmonic potential} \label{subsec:Anharmonic-effects}

Even in the anharmonic case, it is still true that the sloshing mode is the first to be excited from a rapid shift of the trap. Hence, nullifying the sloshing amplitude will provide us with a trajectory very close to optimum. According to Eq.\,\eqref{eq:harmonic_with_corrections}, the anharmonicity can be incorporated 
into an effective harmonic frequency $\omega_{0}^{\prime}=\omega_{0}\left(1-4\frac{\left\langle r^2\right\rangle}{\sigma^2}-\frac{\left\langle\left(z-z_{\cup}\right)^{2}\right\rangle}{z_R^2}\right)$. In the case of a Gaussian beam, the frequency decreases with increasing temperature, an effect referred to as ``softening''. To study how anharmonicity affects the STA, we repeated the experiments of Fig.\,\ref{fig:harmonic_correction}(a) with temperature increased by $\times1.6$ using parametric excitation \cite{Grimm200095}. The axial \emph{in situ} variance of the atomic cloud density $\left\langle\left(z-z_\cup\right)^{2}\right\rangle$ before the transport is about $229\thinspace\unit{\mu m}$ and $272\thinspace\unit{\mu m}$ for the cold and hot clouds, respectively. A good estimate for the radial variance $\left\langle r^{2} \right\rangle$ can be obtained using the known aspect ratio of the trapping frequencies in the axial and radial directions. Using this data together with Eq.\,\eqref{eq:harmonic_with_corrections}, we calculate the ratio between the effective harmonic frequencies in the two experimental conditions and obtain $\frac{\omega_{\mathrm{cold}}}{\omega_{\mathrm{hot}}}=1.027$. For this, we can numerically find a new optimum value for correction amplitude $A_0$. The correction phase $\phi_0$, however, is unaffected by this variation of the effective frequency. 

\begin{figure}
	\centering
	\includegraphics{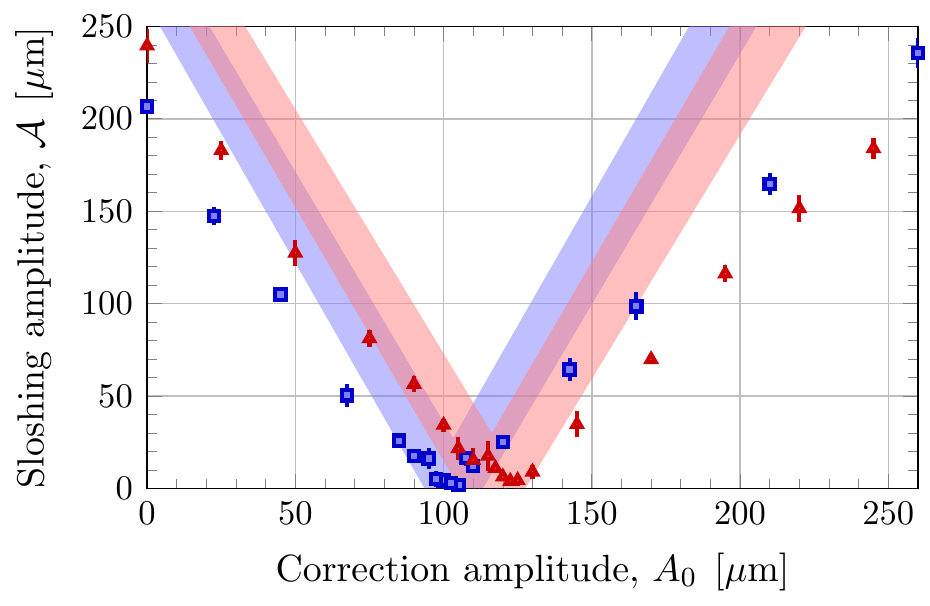}
	\caption{The effect of non-harmonicity in a non-adiabatic transfer. Shown are the sloshing mode amplitudes at $t_f$ versus the harmonic correction amplitude, $A_0$, for cooler (blue squares) and warmer (red triangles) atomic ensembles. As the temperature increases, the atoms experience more the non-harmonic terms and the effective harmonic trapping frequency decreases. The blue data is the same as in Fig.\,\ref{fig:harmonic_correction}(a). The temperature and number of atoms are $T=320\left(30\right)\thinspace\unit{nK}$ and $N=390(20)\cdot {10}^{3}$ for the blue squares, and 		$T=510\left(70\right)\thinspace\unit{nK}$ and $N=380(20)\cdot {10}^{3}$ for the red triangles. Theoretical harmonic calculations appear as ribbons with matching colors, where the width is determined by the frequency uncertainty. The sloshing measurements were performed following a $1.29\thinspace\unit{mm}$ transport lasting $t_{f}=186\thinspace\unit{ms}$ in a trap of axial frequency $\omega_{0}^{\prime}=2\pi\cdot7.16\left(15\right)\thinspace\unit{Hz}$ ($2\pi\cdot6.98\left(15\right)\thinspace\unit{Hz}$) for the cooler (warmer) atoms.
	% data from 16mar18 warm 22apr18 , ribbons freqs are 7.1622(pm0.1469) from the fitting results of cold experiment.
	Correction parameters for both these measurements are $\omega_{0}=2\pi\cdot7.11\thinspace\unit{Hz}$ and $\phi_{0}=279^{\circ}$.
	The frequency shift for the red ribbon was calculated according to the measured extent of both atomic clouds in the trap before the motion with no fitting parameter.
	\label{fig:warm_correction}}
\end{figure}

In Fig.\,\ref{fig:warm_correction} we present the measured sloshing amplitudes for both hot (red triangles) and cold (blue squares) clouds as function of $A_0$ together with theoretical calculations based on $\frac{\omega_{\mathrm{cold}}}{\omega_{\mathrm{hot}}}$ and Eq.\,\eqref{eq:classical_A} (ribbons in matching colors). We find that using the same correction frequency and phase we get a clear minimum with respect to the correction amplitude also for the warmer case, correlative to the calculated amplitudes for $\omega_{0}^{\prime}=2\pi\cdot7.16\left(15\right)\thinspace\unit{Hz}$ and $\omega_{0}^{\prime}=2\pi\cdot6.98\left(15\right)\thinspace\unit{Hz}$ for the cooler and warmer cases, respectively, as calculated from the extents of the clouds before the experiment. In addition, the sloshing frequencies as obtained directly from the measured data following the transport agree with this shifted value of $\omega_{0}^{\prime}$. This demonstrates that the spectral correction technique can account for non-harmonic terms by using an effective trapping frequency.

\subsection{Method II: Septic polynomial trajectory\label{subsec:Septic-polynomial}}

\begin{figure}
\centering
\includegraphics{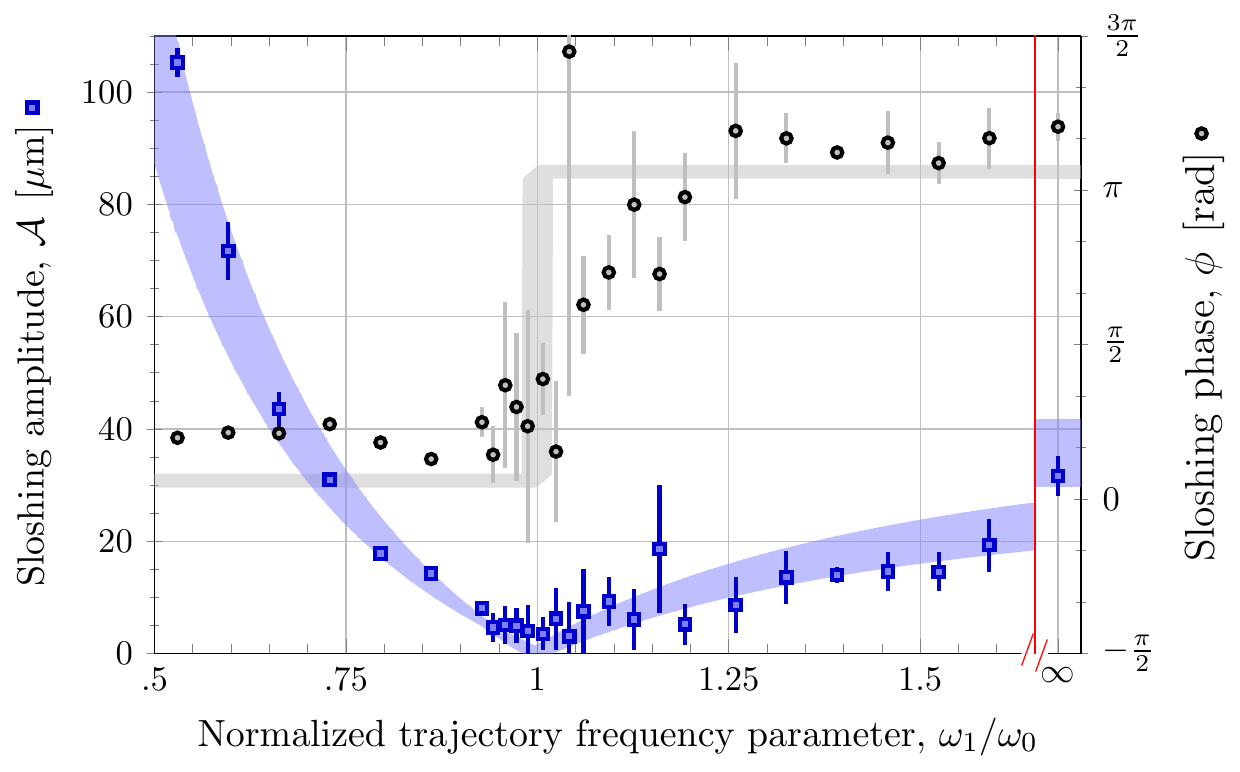}
\caption{Non-adiabatic transfer with a septic polynomial trajectory. Plotted are the sloshing amplitude (blue squares) and phase (gray circles) following a septic trajectory: $z_{\cup}\left(t\right)=z_{7}\left(t\right)+\ddot{z}_{7}\left(t\right)/\omega_{1}^{2}$ versus the frequency parameter $\omega_1$. Theoretical harmonic calculations appear as ribbons with matching colors, where the width is determined by the frequency uncertainty.  Minimum sloshing consistent with zero appears in the vicinity of the real oscillation frequency $\omega_1=\omega_0$. The sloshing measurements performed following a $1.29\thinspace\unit{mm}$ transport during $t_{f}=273\thinspace\unit{ms}$ in a trap of axial frequency $\omega_{0}=2\pi\cdot7.55\left(8\right)\thinspace\unit{Hz}$ ($f_{0}\cdot t_{f}\approx2$).
% data from 2018_03_23, freq is 7.54(8)Hz from sloshing fittings
\label{fig:Septic_trajectory}}
\end{figure}

In the second method, in order to comply with all of the eight boundary conditions, we use a polynomial trajectory of the seventh order, written in terms of the normalized time as: 
\begin{equation}
z_{7}\left(t\right) = d \left[35\left(\frac{t}{t_{f}}\right)^{4}-84\left(\frac{t}{t_{f}}\right)^{5}+70\left(\frac{t}{t_{f}}\right)^{6}-20\left(\frac{t}{t_{f}}\right)^{7}\right]\;.
\label{eq:Septic_trajectory}
\end{equation}
This path for the atoms respects the invariant necessitated boundary conditions and its associated trap trajectory results with zero velocities at motion ends, so it is feasible to be implemented experimentally.
The trap trajectory is then given by Eq.\,\eqref{eq:f_equals_ma}, $z_{\cup}\left(t\right)=z_{7}\left(t\right)+\ddot{z}_{7}\left(t\right)/\omega_{0}^{2}$.
Due to this dependency of the desirable path on the trapping frequency $\omega_{0}$, one is required to provide the later with great accuracy in order to respect the boundary conditions and accomplish the transport with zero sloshing amplitude. In Fig.\,\ref{fig:Septic_trajectory} we present the sloshing amplitudes following such a trajectory where we scan the value of the frequency parameter. Indeed, for the correct value of the frequency parameter we observe a sloshing amplitude consistent with zero. The right-hand side data points are in fact the case where we used the trajectory $z_{7}\left(t\right)$ directly for the trap itself, which results in a considerable excess energy.

\section{Discussion and outlook\label{sec:Discussion}}

While trying to implement previously suggested non-adiabatic trajectories in optical transfer, we have found that in most cases they leave excess energy in the cloud, which we measure as sloshing of the center-of-mass after the trap has stopped. We have identified the source for this behavior in a gap between the required boundary conditions for the trap and realistic constraints. In reality, the trap holding the atoms is at rest before and after the movement. Formerly proposed STA trajectories, however, require the trap to have some initial and final velocity while the atoms are at rest. In principle, it is physically possible, for example by working with two precisely synchronized traps, one holding the atoms at rest and the other moving, then switching between them abruptly when they exactly spatially overlap as the motion commences. Clearly, it is not a practical solution. Hence, we introduced two new boundary conditions requiring the trap to be at rest before and after the motion. We also presented two methods to construct STA trajectories that comply with the new boundary conditions by either correcting an existing trajectory with an added spectral component at the trap frequency or by constructing a new trajectory with a higher order polynomial. Both techniques have demonstrated experimentally in the non-adiabatic regime where the transfer duration is on the order of the inverse trapping frequency. We have also shown that our technique can account for non-harmonic terms in the trapping potential. Thus, our approach is both realistic and flexible, and we anticipate it will be useful in the wide range of applications that can benefit from STA.

\emph{Acknowledgement}. We thank J. Nemirosky for insightful discussions. This research was supported by the Israel Science Foundation (ISF) grant No. 888418, by the United States-Israel Binational Science Foundation (BSF), Jerusalem, Israel, grant No. 2014386, and by the Pazy Research Foundation.

\section*{References}

%\bibliographystyle{unsrt}
%\bibliography{bibliography}

\end{document}